\documentclass[twocolumn,prd,aps,nofootinbib,showpacs]{revtex4}
\usepackage{multirow}
\usepackage{graphicx}
\usepackage[dvips]{color}
\usepackage{amssymb,amsmath}

\begin{document}

\title{Preliminary lattice study of $\sigma$ meson decay width}

\author{Ziwen Fu}
\email{fuziwen@scu.edu.cn}
\affiliation{
Key Laboratory of Radiation Physics and Technology {\rm (Sichuan University)},
Ministry of Education;\\
Institute of Nuclear Science and Technology, Sichuan University,
29 Wangjiang Road, Chengdu, P. R. China
}

\begin{abstract}
We report an exploratory lattice investigation of
$\sigma$ meson decay width using $s$-wave scattering phase
for isospin $I=0$ pion-pion ($\pi\pi$) system.
Rummukainen-Gottlieb formula is used to estimate the scattering phase,
which demonstrate the presence of a resonance around $\sigma$ meson.
Using the effective range formula
we extract the effective $\sigma \to \pi\pi$ coupling constant
as $g_{\sigma \pi\pi} = 2.69(44)$ GeV,
which is consistent with theoretical predictions.
The estimated decay width is about $236 \pm 49$ MeV.
These simulations are carried out on a $16^3\times48$
MILC gauge configuration with the $N_f=2+1$ flavor
of the ``Asqtad'' improved staggered dynamical sea quarks
at $ m_\pi / m_\sigma \approx 0.414$
and the lattice spacing $a \approx 0.15$ fm.
\end{abstract}

\maketitle

\section{ Introduction }
\label{Sec:Introduction}
It is well-known that $\sigma$ meson is a resonance.
In 2010, Particle Data Group (PDG)
lists the meson $f_0(600)$,
which is usually called $\sigma$ meson $I(J^{PC})=0(0^{++})$,
with mass $400-1200$~MeV and 
width~$600-1000$ MeV~\cite{Nakamura:2010zzi}.
Its existence has been established by
some refinements of the experimental analyses~\cite{Ambrosino:2006hb,
Ablikim:2006bz,Ablikim:2004qna,Muramatsu:2002jp,Aitala:2000xu,Asner:1999kj,Svec:1995xr}
and some phenomenological studies~\cite{Oller:1997ti,Hyodo:2010jp,
Mennessier:2010xg,Caprini:2008fc,Yndurain:2007qm,Caprini:2005zr,
Escribano:2002iv,Giacosa:2007bn,Pelaez:2004xp}.
The Dalitz plot analysis of E791~\cite{Aitala:2000xu}
yields its decay width about $324$~MeV.
Moreover, $\sigma$ meson has been extensively studied
with BES data~\cite{Ablikim:2006bz,Ablikim:2004qna},
and most recent analysis gives its pole position: $\sigma$ $(541\pm39) + i(252 \pm42)$ MeV~\cite{Ablikim:2006bz}.

The direct determination of $\sigma$ resonance parameters
from QCD is quite difficult because it is a nonperturbative problem.
However, some theoretical efforts are still taken to investigate $\sigma$ meson
and estimate its resonance parameters~\cite{Oller:1997ti,
Hyodo:2010jp,Mennessier:2010xg,Caprini:2008fc,Yndurain:2007qm,Caprini:2005zr,
Escribano:2002iv,Giacosa:2007bn,Pelaez:2004xp}.
$\sigma$ meson was originally introduced to fit experimental data and
its mass was chosen to reproduce the experimental results.
There is a wide variety for defining its mass and width.
Some authors use the pole approach with the mass and width of resonance
taken from the position of the pole of the T-matrix~\cite{Escribano:2002iv}.
Another way to study the mass and width of resonances is
through the use of the spectral function~\cite{Giacosa:2007bn}, etc.

The most practicable method to nonperturbatively get $\sigma$ resonance parameters
from first principles is using lattice QCD.
To date, there is no report about lattice study
on $\sigma$ resonance parameters, mainly because the reliable calculation
of the rectangular and vacuum diagrams are extremely difficult.
Encouraged by  J.~Nebreda and J.~Pelaez's theoretical investigations
on $\sigma$ resonance~\cite{Nebreda:2010wv},
our previous studies on $\sigma$ mass~\cite{Fu:2011zzh},
$\pi\pi$ scattering length~\cite{Fu:2011bz},
$\pi K$ scattering length~\cite{Fu:2011wc},
and $\kappa$ meson decay width~\cite{Fu:2011xw,Fu:2011xz},
here we explore $\sigma$ resonance parameters through lattice simulation.

In the current study, we extract $\sigma$ decay width
using $s$-wave scattering phase shift of the  $\pi\pi$ system
for isospin $I=0$ channel in the moving frame (MF).
The simulations are performed on a MILC lattice ensemble
with the $2+1$ flavors of the Asqtad improved staggered
dynamical sea quarks~\cite{Bernard:2010fr,Bazavov:2009bb}.
The meson masses extracted from our previous studies~\cite{Fu:2011zzh}
gave $m_\pi/m_\sigma \approx 0.414$,
and lattice parameters were determined by MILC Collaboration.

\section{Methods}
\label{Sec:Methods}
\subsection{Scattering phase}
\label{SubSec:Scattering_Phase}
The $\sigma$  resonance decays into a pair of pions in the $s$-wave.
In the elastic $\pi\pi$ scattering,
the relativistic Breit-Wigner formula (RBWF)~\cite{Nakamura:2010zzi}
can be expressed as
\begin{equation}
\label{eq:BW}
\tan\delta_0=\frac{\sqrt{s} \, \Gamma_R(s)}{m_\sigma^2-s},
\quad s=E_{CM}^2 .
\end{equation}
where  $E_{CM}$ is its center-of-mass (CM) energy of $\pi\pi$ system,
$\delta_0$ is its $s$-wave scattering phase,
and decay width $\Gamma_R(s)$ can be written by means of
the effective $\sigma\rightarrow\pi\pi$ coupling constant
$g_{\sigma \pi\pi}$~\cite{Nebreda:2010wv},
\begin{equation}
\label{eq:g_kpK_formula}
\Gamma_R(s)=\frac{g^2_{\sigma\pi\pi}}{8\pi}\frac{p}{s} ,
\quad p = \sqrt{\frac{s}{4} - m_\pi^2 } .
\end{equation}
By inspecting eqs.~(\ref{eq:BW}) and (\ref{eq:g_kpK_formula}),
an expression of the $s$-wave scattering phase
with the invariant mass $\sqrt{s}$ is given by so-called effective range formula (ERF), namely,
\begin{equation}
\label{eq:effective_range_formula}
\tan{\delta_0}=\frac{g^2_{\sigma\pi\pi}}{8\pi}
\frac{p}{ \sqrt{s}(m_\sigma^2- s)} ,
\end{equation}
which enable us  either a linear fit or solving 
for two unknown parameters: $g_{\sigma\pi\pi}$ and $m_\sigma$.
Then $\sigma$ decay width $\Gamma_\sigma$ can be estimated through
\begin{equation}
\label{eq:decay_width}
\Gamma_\sigma = \Gamma_R(s)\Bigg|_{s = m_\sigma^2} =
\frac{g^2_{\sigma\pi\pi}}{8\pi}\frac{p_\sigma}{m_\sigma^2} ,
\quad p_\sigma =
\sqrt{ \frac{m_\sigma^2}{4} - m_\pi^2} .
\end{equation}
Thus, equations~(\ref{eq:effective_range_formula}) and (\ref{eq:decay_width})
give us a way to obtain $m_\sigma$ and $\Gamma_\sigma$
through lattice simulation.

\subsection{Finite size formula}
In this work, we focus on $\sigma$ meson decay
into a pair of pions in the $s$-wave,
and only investigate the $\pi\pi$ system  with isospin $(I,I_z)=(0,0)$.

\subsubsection{Center of mass frame}
In the non-interacting case,
the energy eigenvalues of $\pi\pi$ system are given by
$$
E = 2 \sqrt{m_\pi^2+ p^2} ,
$$
where $p=|{\mathbf p}|,  {\mathbf p}=(2\pi/L){\mathbf n}$, and
${\mathbf n}\in \mathbb{Z}^3$.
For a typical lattice study, the energy for ${\mathbf n} \ne 0$
is much larger than sigma mass $m_\sigma$.
For our case, the lowest energy for ${\mathbf n} \ne 0$
calculated from $m_\pi$ and $m_\sigma$
is $E = 1.56 \times m_\sigma$,
which is obviously not eligible to study $\sigma$ meson.
Thus, we can only consider ${\mathbf n} = 0$ case,
and the energy $E = 0.828 \times m_\sigma$,
which is still a not good choice.

In the interacting case,
the energy eigenvalues are moved by the hadronic interaction
from $E$ to $\overline{E}$,
and the energy eigenvalue for the $\pi\pi$ system is given by
$$
\overline{E} = 2\sqrt{m_\pi^2 + k^2} ,
\quad k = \frac{2\pi}{L}q,
$$
where $q$ is not required to be an integer.
In the CM frame these energy eigenvalues
transform  under the irreducible representation
$\Gamma = T_1^+$ of the cubic group $O_h$.
It is the famous L\"uscher formula that relates
the energy $\overline{E}$
to the scattering phase  $\delta_0$~\cite{Luscher:1990ux,Luscher:1990ck}, namely,
\begin{eqnarray}
\label{eq:CMF}
\tan\delta_0(k) &=& \frac{\pi^{3/2}q}{\mathcal{Z}_{00}(1;q^2)}, \cr
\mathcal{Z}_{00}(s;q^2) &=& \frac{1}{\sqrt{4\pi}}
\sum_{{\mathbf n}\in\mathbb{Z}^3}
\frac{1}{\left(|{\mathbf n}|^2-q^2\right)^s} .
\end{eqnarray}
The zeta function can be efficiently evaluated by the method
in ref.~\cite{Yamazaki:2004qb}.

\subsubsection{Moving frame}
To make the energy of the $\pi\pi$ system is closer to sigma mass $m_\sigma$,
we adopt the moving frame (MF)~\cite{Rummukainen:1995vs}
with total momentum ${\mathbf P}=(2\pi/L){\mathbf d}$,
${\mathbf d}\in\mathbb{Z}^3$.
For the non-interacting case its energy eigenstates are given by
$$
E_{MF} = \sqrt{m_\pi^2+p_1^2} + \sqrt{m_\pi^2+ p_2^2} ,
$$
where $p_1=|{\mathbf p}_1|$,$p_2=|{\mathbf p}_2|$,
and ${\mathbf p}_1$, ${\mathbf p}_2$ define
three-momenta of two pions, respectively, which meet
\begin{equation}
\label{eq:sum_boundary}
{\mathbf p}_1=\frac{2\pi}{L}{\mathbf n}_1 , \quad
{\mathbf p}_2=\frac{2\pi}{L}{\mathbf n}_2 , \quad
{\mathbf n}_1,{\mathbf n}_2\in \mathbb{Z}^3, \quad
{\mathbf P} = {\mathbf p}_1 + {\mathbf p}_2 .
\end{equation}
In the MF, the CM is moving
with a velocity of ${\mathbf v}={\mathbf P}/E_{MF}$.
Using the Lorentz transformation
with a boost factor $\gamma=1/\sqrt{1-{\mathbf v}^2}$,
the $E_{CM}$ can be calculated by
\begin{equation}
E_{CM} = \gamma^{-1}E_{MF} = 2\sqrt{m_\pi^2 + p^{*2}}  ,
\end{equation}
where
\begin{equation}
\label{eq:Energy_pp}
p^*=| {\mathbf p}^*|\,,\quad
{\mathbf p}^*={\mathbf p}^*_1=-{\mathbf p}^*_2 =
\frac{1}{2} \vec{\gamma}^{-1} ({\mathbf p}_1 -{\mathbf p}_2 ),
\end{equation}
here and hereafter we denote the CM  momenta
with an asterisk $(^\ast)$,
the boost factor acts in the direction of ${\mathbf v}$, and
we adopt the shorthand notation
\begin{equation}
\vec{\gamma}^{-1}{\mathbf p} =
\gamma^{-1}{\mathbf p}_{\parallel}+{\mathbf p}_{\perp},
\end{equation}
where ${\mathbf p}_{\parallel}$ and ${\mathbf p}_{\perp}$ are
parallel and perpendicular components of ${\mathbf p}$.
We note that the ${\mathbf p}^*$ are quantized as
\begin{equation}
{\mathbf p}^* =\frac{2\pi}{L}{\mathbf r}\,, \quad
{\mathbf r} \in P_{\mathbf d} ,
\end{equation}
where
\begin{equation}
\label{eq:set_Pd_MF}
P_{\mathbf d} = \left\{ {\mathbf r} \left|  {\mathbf r} = \vec{\gamma}^{-1}
\left({\mathbf n}+\frac{1}{2}{\mathbf d} \right), \right. \ {\mathbf n}\in\mathbb{Z}^3 \right\} .
\end{equation}
We are specially interest in one MF,
which are one pion at rest, one pion with momentum
${\mathbf p} = (2\pi / L) {\mathbf e}_3$ (${\mathbf d} = {\mathbf e}_3$)
and $\sigma$ meson with momentum  ${\mathbf P} = {\mathbf p}$.
For our case, we found that its invariant mass
takes $\sqrt{s} = 0.994\times m_\sigma$,
which is significantly closer to $m_\sigma$ than that in the CM frame.
Thus, in this work we will only study this case.

In the interacting case, the $\overline{E}_{CM}$ is given by
\begin{equation}
\label{eq:continuum_dis_rel}
\overline{E}_{CM} = 2 \sqrt{m_\pi^2 + k^{2}} ,
\quad k = \frac{2\pi}{L} q .
\end{equation}
where $q$ is no longer an integer.
In this  work, we only use one MF with ${\mathbf d}={\mathbf e}_3$,
where the energy eigenstates transform
under the irreducible representation $A_2^-$ of the tetragonal group $D_{4h}$.
We use the famous Rummukainen-Gottlieb formula to relate the energy eigenstates
to the $\pi\pi$ scattering phase shift $\delta_0$, namely,
\begin{equation}
\label{eq:Luscher_MF}
\tan\delta_0(k)=
\frac{\gamma\pi^{3/2}q}{\mathcal{Z}_{00}^{\mathbf d}(1;q^2)} ,
\end{equation}
where we suppose that the higher phase shifts $\delta_l$  ($l = 2, 4, 6,. . .$)
are negligible, and the zeta function
\begin{equation}
Z_{00}^{{\mathbf d}} (s; q^2) = \sum_{{\mathbf r}\in P_{\rm d}}
\frac{1} { ( |{\mathbf r}|^2 - q^2)^s } ,
\label{zetafunction_MF}
\end{equation}
here the set $P_{\mathbf d}$  is defined in eq.~(\ref{eq:set_Pd_MF}).
$k$ is the scattering momentum defined through the invariant mass $\sqrt{s}$, namely,
$\sqrt{s} =  \sqrt{ E_{MF}^2 - p^2 } = 2 \sqrt{ m_\pi^2 + k^2}$.
The calculation method of $Z_{00}^{{\mathbf d}} (1; q^2)$
is discussed in refs.~\cite{Yamazaki:2004qb,Fu:2011xw}.

\subsection{Correlation matrix}
\label{SubSec:Correlation_matrix }
To evaluate the energy eigenstates,
we build a matrix of the correlation function, namely,
\begin{equation}
C(t) = \left(
\begin{array}{ll}
\langle 0 | {\cal O}_{\pi\pi}^\dag(t)  {\cal O}_{\pi\pi}(0) | 0 \rangle &
\langle 0 | {\cal O}_{\pi\pi}^\dag(t)  {\cal O}_\sigma(0)   | 0 \rangle
\vspace{0.3cm}\\
\langle 0 | {\cal O}_\sigma^\dag(t)   {\cal O}_{\pi\pi}(0)  | 0 \rangle &
\langle 0 | {\cal O}_\sigma^\dag(t)   {\cal O}_\sigma(0)    | 0 \rangle
\end{array} \right) ,
\label{eq:CorrMat}
\end{equation}
where ${\cal O}_\sigma(t)$ and ${\cal O}_{\pi\pi}(t)$ are
interpolating operators for $\sigma$ meson and $\pi\pi$ system, respectively.
These interpolating operators employed here are
the same as in our previous studies~\cite{Bernard:2007qf,Fu:2011zzh,Fu:2011zzl}.
However, to make this work self-contained,
we will give the necessary definitions below.

\subsubsection{$\pi\pi$ sector}
Let us study $\pi\pi$ scattering of two Nambu-Goldstone
pions in the Asqtad-improved staggered dynamical fermion formalism.
Here we follow original derivations and notations
in refs.~\cite{Sharpe:1992pp,Kuramashi:1993ka,Fukugita:1994ve}.
Using operators ${\cal O}_\pi(x_1)$, ${\cal O}_\pi(x_2)$,
${\cal O}_\pi(x_3)$, ${\cal O}_\pi(x_4)$ for pions at points $x_1$, $x_2$,
$x_3$ and $x_4$, respectively,
with pion interpolating operators
\begin{eqnarray}
 \pi^+(t) &=&
 -\sum_{\mathbf{x}}
 \bar{d}(\mathbf{x}, t) \gamma_5 u(\mathbf{x}, t), \cr
 \pi^-(t) &=&
 \sum_{\mathbf{x}}
 \bar{u}(\mathbf{x},t) \gamma_5 d(\mathbf{x},t), \cr
 \pi^0(t) &=&
 \frac{1}{\sqrt{2}}\sum_{\mathbf{x}} [
 \bar{u}(\mathbf{x},t) \gamma_5 u(\mathbf{x},t) -
 \bar{u}(\mathbf{x},t) \gamma_5 d(\mathbf{x},t) ], \nonumber
\end{eqnarray}
we express the $\pi\pi$ four-point functions as
$$
C_{\pi\pi}(x_4,x_3,x_2,x_1) =
\bigl< {\cal O}_\pi(x_4) {\cal O}_{\pi}(x_3)
{\cal O}_\pi^{\dag}(x_2) {\cal O}_{\pi}^{\dag}(x_1)\bigr> .
$$

After summing over the spatial coordinates,
we achieve the $\pi\pi$ four-point function with the momentum ${\mathbf p}$,
\begin{eqnarray}
\label{EQ:4point_pK_mom}
C_{\pi\pi}({\mathbf p}, t_4,t_3,t_2,t_1) &=&
\sum_{\mathbf{x}_1}\sum_{\mathbf{x}_2}\sum_{\mathbf{x}_3}\sum_{\mathbf{x}_4}
e^{ i{\mathbf p} \cdot ({\mathbf{x}}_4 -{\mathbf{x}}_2) } \cr
&&\times C_{\pi\pi}(x_4,x_3,x_2,x_1) ,
\end{eqnarray}
where $x_1 \equiv (\mathbf{x}_1, t_1)$,
      $x_2 \equiv (\mathbf{x}_2, t_2)$,
      $x_3 \equiv (\mathbf{x}_3, t_3)$,
      $x_4 \equiv (\mathbf{x}_4, t_4)$, and
      $t\equiv t_3 - t_1$.

\begin{figure}[h]
\begin{center}
\includegraphics[width=7.5cm]{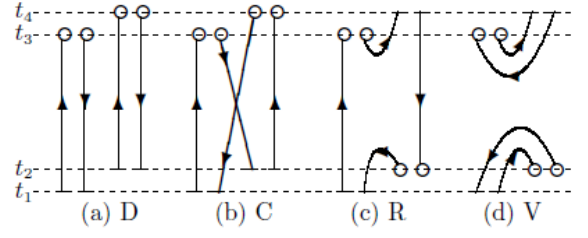}
\end{center}
\vspace{-0.5cm}
\caption{ \label{fig:diagram}
Diagrams contributing to $\pi\pi$ four-point functions.
Short bars stand for wall sources.
Open circles are sinks for local pion operators.
}
\end{figure}

To avert the Fierz rearrangement of
the quark lines~\cite{Kuramashi:1993ka,Fukugita:1994ve},
we choose  $t_1 =0, t_2=1, t_3=t$, and $t_4 = t+1$.
We then construct the $\pi\pi$ operators for the $I =0$ channel as
\begin{eqnarray}
\label{EQ:op_pipi}
 {\cal O}_{\pi\pi}^{I=0} (\mathbf{p},t) &=&
 \frac{1}{\sqrt{3}}
 \Bigl\{  \pi^{-}(t) \pi^{+}(\mathbf{p},t+1) +
          \pi^{+}(t) \pi^{-}(\mathbf{p},t+1)  \cr
 && -\pi^{0}(t) \pi^{0}(\mathbf{p},t+1) \Bigl\} .
\end{eqnarray}
These operators belong to $A_2^-$ and the $(I,I_z)=(0, 0)$.

In the isospin limit, only four diagrams contribute to $\pi\pi$ scattering amplitudes~\cite{Sharpe:1992pp,Kuramashi:1993ka,Fukugita:1994ve},
which are plotted in figure~\ref{fig:diagram},
and labeled as direct($D$), crossed ($C$), rectangular ($R$),
and vacuum ($V$) diagrams, respectively.
It is well-known that the reliable evaluation of
the rectangular ($R$) and vacuum diagrams ($V$) are extremely difficult.
we tackle it by evaluating $T$ quark propagators~\cite{Kuramashi:1993ka,Fukugita:1994ve,Fu:2011bz},
namely, each propagator, which corresponds to a moving wall source
at $ t = 0, \cdots, T-1$, are denoted by
$$
\sum_{n''}D_{n',n''}G_t(n'') = \sum_{\mathbf{x}}
\delta_{n',({\mathbf{x}},t)}, \quad 0 \leq t \leq T-1 .
$$
The combination of $G_t(n)$ that we apply for $\pi\pi$ four-point
functions is shown in figure~\ref{fig:diagram}.
For the non-zero momentum,  we used an up quark source with $1$, and an
anti-up quark source with $e^{i{\mathbf p} \cdot {\mathbf{x}} }$
(except for $V$, where we use $1$)
on each site for two pion creation operator, respectively.
$D$, $C$, $R$ and $V$ are schematically shown in figure~\ref{fig:diagram},
and we can express them by means of the quark propagators $G$, namely,
\begin{widetext}
\begin{eqnarray}
\label{eq:dcr}
\hspace{-1.0cm}C^D_{\pi\pi}({\mathbf p},t_4,t_3,t_2,t_1) \hspace{-0.1cm}&= &\hspace{-0.1cm}
\mbox{Re} \hspace{-0.1cm}\sum_{ \mathbf{x}_3, \mathbf{x}_4}
e^{ i{\mathbf p} \cdot {\mathbf{x}}_4 }
\left\langle \mbox{Tr}
[G_{t_1}^{\dag}({\mathbf{x}}_3, t_3) G_{t_1}({\mathbf{x}}_3, t_3)
 G_{t_2}^{\dag}({\mathbf{x}}_4, t_4) G_{t_2}({\mathbf{x}}_4, t_4) ] \right\rangle,\cr
\hspace{-1.0cm}C^C_{\pi\pi}({\mathbf p},t_4,t_3,t_2,t_1) \hspace{-0.1cm}&=&\hspace{-0.1cm}
\mbox{Re} \hspace{-0.1cm} \sum_{\mathbf{x}_3, \mathbf{x}_4}
e^{ i{\mathbf p} \cdot {\mathbf{x}}_4 }
\left\langle \mbox{Tr}
[G_{t_1}^{\dag}({\mathbf{x}}_3, t_3) G_{t_2}({\mathbf{x}}_3, t_3)
 G_{t_2}^{\dag}({\mathbf{x}}_4, t_4) G_{t_1}({\mathbf{x}}_4, t_4) ] \right\rangle,\cr
\hspace{-1.0cm}C^R_{\pi\pi}({\mathbf p}, t_4,t_3,t_2,t_1) \hspace{-0.1cm}&=& \hspace{-0.1cm}
\mbox{Re} \hspace{-0.1cm} \sum_{\mathbf{x}_2,\mathbf{x}_3}
e^{ i{\mathbf p} \cdot {\mathbf{x}}_2 }
\left\langle \mbox{Tr}
[G_{t_1}^{\dag}({\mathbf{x}}_2, t_2) G_{t_4}({\mathbf{x}}_2, t_2)
 G_{t_4}^{\dag}({\mathbf{x}}_3, t_3) G_{t_1}({\mathbf{x}}_3, t_3) ] \right\rangle, \cr
\hspace{-1.0cm}C_{\pi\pi}^V({\mathbf p},t_4,t_3,t_2,t_1) \hspace{-0.1cm}&=&\hspace{-0.1cm}
\mbox{Re} \hspace{-0.1cm} \sum_{\mathbf{x}_2,\mathbf{x}_3 }
\hspace{-0.15cm}e^{ i{\mathbf p} \cdot ({\mathbf{x}}_2  - {\mathbf{x}}_3) }
\hspace{-0.1cm}\left\langle \mbox{Tr}
[G_{t_1}^{\dag}\hspace{-0.05cm}(\mathbf{x}_2, t_2)
G_{t_1}\hspace{-0.05cm}(\mathbf{x}_2, t_2)
G_{t_4}^{\dag}\hspace{-0.05cm}(\mathbf{x}_3, t_3) G_{t_4}\hspace{-0.05cm}(\mathbf{x}_3, t_3) ]
\right\rangle \hspace{-0.1cm} .
\end{eqnarray}
\end{widetext}
From our previous studies~\cite{Fu:2011zzh,Bernard:2007qf,Fu:2011zzl},
we found that when we calculate the disconnected part of
the sigma correlator with non-zero momenta,
the subtraction of the vacuum expectation value is not needed.
Similarly, the vacuum diagram here is not accompanied
by a vacuum subtraction for the cases with non-zero momenta.

As discussed in refs.~\cite{Kuramashi:1993ka,Fukugita:1994ve},
the rectangular and vacuum diagrams create gauge-variant noise,
which are reduced by performing the gauge field average without gauge fixing in this work.
All four diagrams in figure~\ref{fig:diagram} are required
to study $\pi\pi$ scattering in the $I=0$ channel.
As investigated in ref.~\cite{Kuramashi:1993ka,Fukugita:1994ve},
in the isospin limit, the $\pi\pi$ correlator
for the $I=0$ channel can be written 
with the combinations of four diagrams, namely,
\begin{equation}
\label{EQ:phy_I12_32}
C_{\pi\pi}(t)  \equiv
\left\langle {\cal O}_{\pi\pi}(t) | {\cal O}_{\pi\pi}(0) \right\rangle =
D + \frac{1}{2} C - 3 R + \frac{3}{2}V  ,
\end{equation}
where the operator ${\cal O}_{\pi\pi}$ denoted in eq.~(\ref{EQ:op_pipi})
creates a $\pi\pi$ state with the total isospin $0$.
In practice we also evaluate the ratios
\begin{equation}
\label{EQ:ratio}
R^X(t) = \frac{ C_{\pi\pi}^X({\mathbf p},0,1,t,t+1) }
{ C_\pi ({\mathbf 0},0,t) C_\pi({\mathbf p},1,t+1) },
\hspace{0.25cm} X\hspace{-0.05cm}=D, C, R, \ {\rm and} \ V ,
\end{equation}
where $C_\pi ({\mathbf 0}, 0,t)$ and $C_\pi ({\mathbf p},1,t+1)$ are
the two-point pion correlators with the momentum
${\mathbf 0 }$ and ${\mathbf p}$, respectively.

We should bear in mind that,
the contributions of non-Nambu-Goldstone
pions in the intermediate states is exponentially
suppressed for large $t$ due to its heavier masses
compared to these of the Nambu-Goldstone
pion~\cite{Sharpe:1992pp,Kuramashi:1993ka,Fukugita:1994ve}.
Hence, we think that $\pi\pi$ interpolator
does not greatly couple to other $\pi\pi$  tastes,
and neglect this systematic errors.

\subsubsection{$\sigma$ sector}
\label{subsub:sigma_sec}
In our previous studies~\cite{Fu:2011zzh,Bernard:2007qf,Fu:2011zzl},
we give a detailed procedure to evaluate
$\langle 0 | \sigma^\dag(t)  \sigma   (0) | 0 \rangle$ .
To simulate the correct number of quark species,
we use an interpolation operator
with the isospin $I=0$ and $J^{P}=0^{+}$ at source and sink,
$$
{\cal O}(x)  \equiv
\sum_{a, g}
\frac{ \bar u^a_g( x ) u^a_g( x ) +
       \bar d^a_g( x ) d^a_g( x ) }{ \sqrt{2n_r} } ,
$$
where $g$   is the index of the taste replica,
      $n_r$ is the number of the taste replicas,
      $a$   is the color index.
After performing the Wick contractions of fermion fields,
and summing over the taste index,
for the light quark Dirac operator $M$,
we obtain the time slice correlator $C(t)$ with the momentum ${\bf p}$
\begin{eqnarray}
\label{EQ:CBB}
C(\mathbf{p},t) \hspace{-0.2cm}&=&\hspace{-0.2cm}
-\frac{1}{2} \sum_{ \mathbf{x} }
e^{i \mathbf{p} \cdot \mathbf{x} }
\langle
\mbox{Tr} M^{-1} (\mathbf{x},t; \mathbf{x},t)
\mbox{Tr} M^{-1} ({\bf 0},0;{\bf 0},0)
\rangle \cr
\hspace{-0.2cm}&+&\hspace{-0.2cm}
\sum_{ \mathbf{x} }(-)^x e^{i \mathbf{p}\cdot \mathbf{x} }
\langle \mbox{Tr}
[ M^{-1}     (\mathbf{x},t;{\bf 0},0)
  M^{-1^\dag} \mathbf{x},t;{\bf 0},0) ]
\rangle, \nonumber
\end{eqnarray}
where the first and second  terms are the
quark-line disconnected and connected contributions,
respectively~\cite{Fu:2011zzh,Bernard:2007qf,Fu:2011zzl}.
For the staggered quarks, the meson propagators behave as
\begin{equation}
\label{sfits:ch7}
{\cal C}(t) =
\sum_i A_i e^{-m_i t} + \sum_i A_i^{\prime}(-1)^t e^{-m_i^\prime t}  +(t \rightarrow N_t-t),
\end{equation}
where the oscillating term is a particle with opposite parity.
For $\sigma$ correlator,
we take only one mass with each parity in eq.~(\ref{sfits:ch7})~\cite{Fu:2011zzh,Bernard:2007qf,Fu:2011zzl}.
Thus, the $\sigma$ correlator was then fit to
\begin{equation}
\label{eq:kfit}
  C_{\sigma}(t)  = b_{\sigma}e^{-m_{\sigma}t} +
  b_{\eta_A}(-)^t e^{-M_{\eta_A}t} + (t \rightarrow N_t-t),
\end{equation}
where $b_{\eta_A}$ and $b_{\sigma}$ are two overlap factors.

\subsubsection{Off-diagonal sector}
To avoid the Fierz rearrangement of the quark lines,
we choose $t_1 =0, t_2=1$, and $t_3=t$
for the $\pi\pi \to \sigma$ three-point function,
and choose $t_1 =0, t_2=t$, and $t_3=t+1$
for the $\sigma \to \pi\pi$ three-point function~\cite{Fukugita:1994ve}.
\begin{figure}[th]
\begin{center}
\includegraphics[width=6cm,clip]{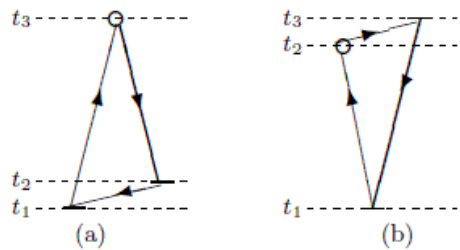}
\end{center}
\vspace{-0.5cm}
\caption{ \label{fig:3diagram}
Diagrams contributing to $\pi\pi \to \sigma$ and $\sigma \to \pi\pi$.
Short bars represent the wall sources.
Open circles stand for the sinks of local pion operator.
(a) Quark line contractions of $\pi\pi \to \sigma$ and
(b) Quark line contractions of $\sigma \to \pi\pi$.
}
\end{figure}
The quark line diagrams contributing to $\pi\pi \to \sigma$
and $\sigma \to \pi\pi$ three-point functions are plotted
in figure~\ref{fig:3diagram}(a) and figure~\ref{fig:3diagram}(b), respectively.
The $\pi\pi \to \sigma$  three-point function can be easily evaluated.
However, the calculation of the $\sigma \to \pi\pi$ three-point function
is quite difficult.
For non-zero momenta, we adopted an up quark source with $1$,
and an anti-up quark source with $e^{i{\mathbf{p}}\cdot{\mathbf{x}}}$
on each site for pion creation operator.
The $\sigma \to \pi\pi$ and $\pi\pi \to \sigma$  three-point functions
are schematically displayed in figure~\ref{fig:diagram},
and we can write them using quark propagators $G$, namely,
\begin{eqnarray}
\label{eq:dcr3}
\hspace{-0.5cm}C_{\pi\pi \to \sigma} ({\mathbf p},t_3,t_2,t_1) &=& \cr
&&\hspace{-3.5cm}\mbox{Re} \sum_{ {\mathbf{x}}_3, {\mathbf{x}}_1}
  e^{ i {\mathbf p} \cdot {\mathbf{x}}_3 }
\left\langle \mbox{Tr}
[G_{t_1}({\mathbf{x}}_3, t_3) G_{t_2}^{\dag}({\mathbf{x}}_3, t_3)
 G_{t_2}^{\dag}({\mathbf{x}}_1, t_1) ] \right\rangle,\cr
\hspace{-0.5cm}C_{\sigma \to \pi\pi}({\mathbf p},t_3,t_2,t_1) &=& \cr
&&\hspace{-3.5cm}\mbox{Re} \sum_{ {\mathbf{x}}_2, {\mathbf{x}}_1}
\hspace{-0.1cm} e^{ i {\mathbf p} \cdot {\mathbf{x}}_2 }
\hspace{-0.1cm}  \left\langle  \mbox{Tr}
[G_{t_1}({\mathbf{x}}_2, t_2)
 G_{t_3}^{\dag}({\mathbf{x}}_2, t_2)
 G_{t_3}^{\dag}({\mathbf{x}}_1, t_1) ] \right\rangle,
\end{eqnarray}

\subsection{ Extraction of energies }
\label{SubSec:Extraction of energies}
To map out ``avoided level crossings''
between $\sigma$ resonance and its decay products,
we separate the ground state from first excited state
by calculating the matrix of correlation function
$C(t)$ in eq.~(\ref{eq:CorrMat}).
To extract two lowest energy eigenvalues,
we utilize the variational method~\cite{Luscher:1990ck}
and build a ratio of correlation function matrices as
\begin{equation}
M(t,t_R) = C(t) \, C^{-1}(t_R) ,
\label{eq:M_def}
\end{equation}
with some reference time slice $t_R$~\cite{Luscher:1990ck}.
The two lowest energy states can be extracted
through a fit to two eigenvalues $\lambda_n (t,t_R)$ ($n=1,2$) of
matrix $M(t,t_R)$.
Since we work on the staggered fermion,
$\lambda_n (t,t_R)$ ($n=1,2$) explicitly has an oscillating term~\cite{Fu:2011xw,Mihaly:1997}, namely,
\begin{eqnarray}
\label{Eq:asy}
\lambda_n (t, t_R) &=&  A_n
\cosh\left[-E_n\left(t-\frac{T}{2}\right)\right] \cr
&&+ (-1)^t B_n
\cosh\left[-E_n^{\prime}\left(t-\frac{T}{2}\right)\right] ,
\end{eqnarray}
for a large $t$, which mean $0 \ll t_R < t \ll T/2$ to
suppress both the excited states and wrap-around contributions~\cite{Feng:2009ij}.
Here we assume $\lambda_1(t,t_R) > \lambda_2(t,t_R)$.

\section{Lattice calculation}
\label{sec:latticeCal}

We used MILC lattice with $2+1$ dynamical flavors of
Asqtad-improved staggered dynamical fermions.
We worked on a $0.15$ fm lattice ensemble
of $360$ $16^3 \times 48$ gauge configurations
with bare quark masses $am_{ud}/am_s = 0.0097/0.0484$
and bare gauge coupling $10/g^2 = 6.572$.
The lattice extent $L$ is about $2.5 {\mathrm fm}$,
the $u$ and $d$ quark masses are degenerate and
the lattice spacing $a^{-1}=1.358^{+35}_{-13}$ GeV~\cite{Bernard:2010fr,Bazavov:2009bb}.

We use the standard conjugate gradient method to obtain
the required matrix element of the inverse fermion matrix.
Periodic boundary condition is imposed to three spatial directions
and temporal direction.
We compute the propagators on all the time slices
for the $\pi\pi$ correlation functions,.
After averaging the correlator over all $48$ possible values,
the statistics are greatly improved
since we can put pion source at all possible time slices.

We calculate the off-diagonal correlator $C_{21}(t)$ by
$$
C_{21}(t) = \frac{1}{T}\sum_{t_s}
\left\langle\sigma(t+t_s)(\pi\pi)^\dag(t_s)\right\rangle ,
$$
where we sum the correlator over all time slices and average it.
Using the relation $C_{12}(t)=C_{21}^\ast(t)$,
we obtain another off-diagonal correlator $C_{12}(t)$
for free~\cite{Aoki:2007rd}.

For the $\sigma$ correlator, $C_{22}(t)$,
we can use the available propagators in~\cite{Fu:2011zzh}
to build the $\sigma$ correlator
$$
C_{22}(t)=\frac{1}{T}\sum_{t_s}\left\langle\sigma^\dag(t+t_s) \sigma(t_s)\right\rangle ,
$$
where, again, we average all the possible correlators.
One thing we must stress is that we use the $Z_2$ noisy estimators
based on the random color fields to measure the disconnected contribution
of the sigma correlator~\cite{Fu:2006uw}.
In our previous work~\cite{Fu:2006uw}, we have presented the detailed procedures
for the implementation of the $Z_2$ method.
Using the standard discussed in ref.~\cite{Muroya:2001yp},
we determine that $1000$ noise $Z_2$ sources are sufficiently reliable
to measure the disconnected part.

We also measure two-point pion correlators, namely,
\begin{eqnarray}
C_\pi(t;{\mathbf 0}) &=&
\langle 0|\pi^\dag ({\mathbf 0},t) \pi({\mathbf 0},0) |0\rangle , \cr
C_\pi(t;{\mathbf p}) &=&
\langle 0|\pi^\dag ({\mathbf p},t) \pi({\mathbf p},0) |0\rangle ,
\label{eq:Gpi}
\end{eqnarray}
where the $C_\pi (t;{\mathbf 0})$ and  $C_\pi (t;{\mathbf p})$ are
correlators for the pion with the momentum
${\mathbf 0 }$ and ${\mathbf p}$, respectively.

\section{Simulation results }
\label{Sec:Results}
\subsection{Time correlation function}
\begin{figure}[th]
\begin{center}
\includegraphics[width=8cm,clip]{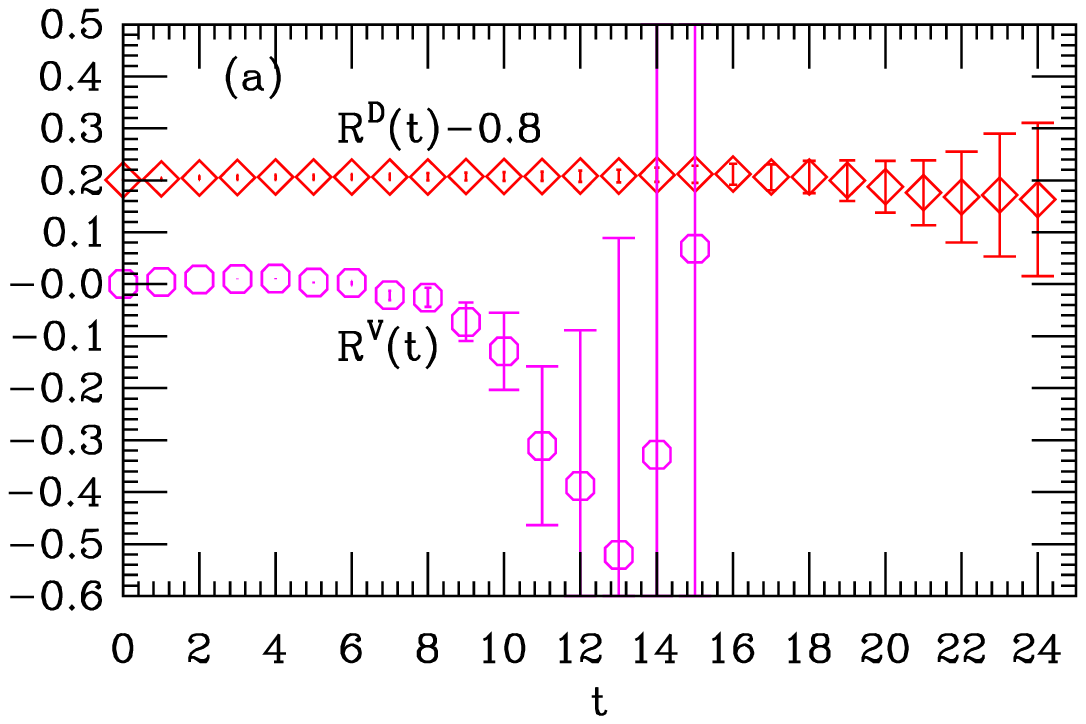}
\includegraphics[width=8cm,clip]{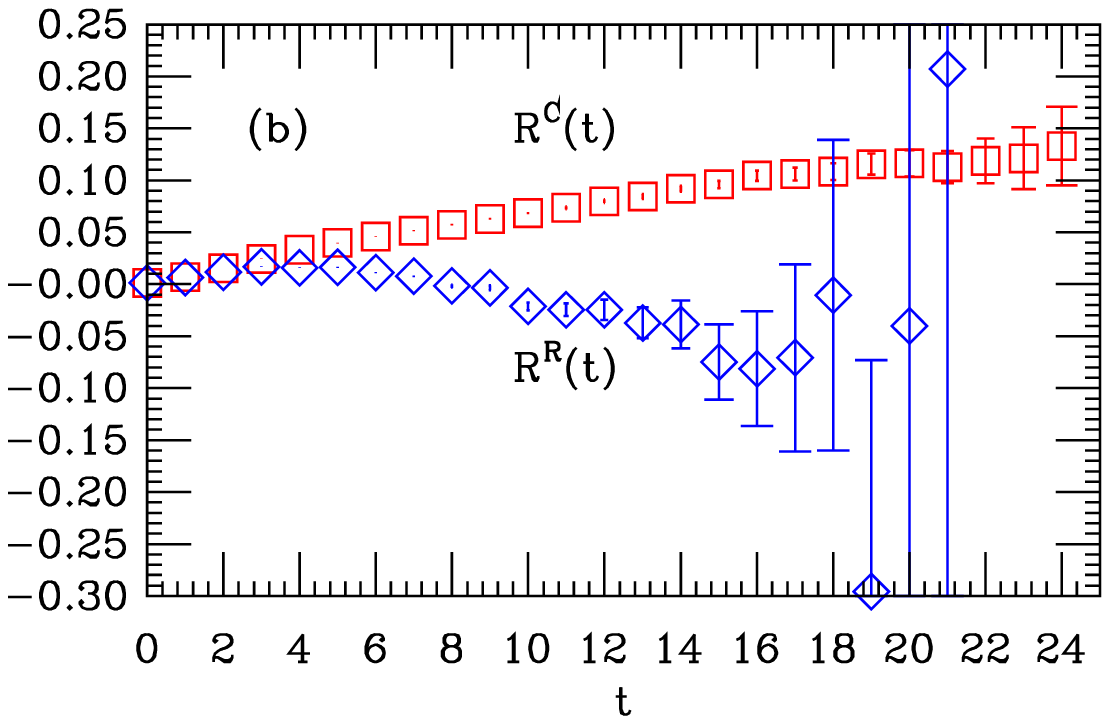}
\end{center}
\vspace{-0.5cm}
\caption{
Individual amplitude ratios $R^X(t)$ for $\pi\pi$ four-point function
as functions of $t$.
(a) Direct diagram shifted by $0.8$ (red diamonds) and vacuum diagram (magenta octagons);
(b) crossed (red squares) and rectangular (blue diamonds) diagrams.
\label{fig:ratio}
}
\end{figure}

In figure~\ref{fig:ratio} the individual ratios,
$R^X$ ($X=D, C$, $R$ and $V$)~\footnote{
We can verify that when $t \ll T/2$,
we can approximately estimate the energy shift $\delta E$ from ratio $R^X$.
}
are plotted as the functions of $t$.
It is quite noisy for the disconnected diagram ($V$),
but we can still get a signal up until time separation $t \sim 14$.
Clear signals observed up until $t \sim 20$ for the rectangular amplitude
and up until $t \sim 14$ for the vacuum amplitude
show that the technique with the moving wall source
without gauge fixing used in this paper is practically applicable.

The values of the direct amplitude $R^D$ is quite close to unity,
indicating a weak interaction in this channel.
The crossed amplitude increases linearly, implying a repulsion in this channel.
After an beginning increase up to $t \sim 4$, the rectangular amplitude
shows a approximately linear decrease up untill $t \sim 15$,
suggesting an attractive force between two pions.
We can note that the crossed and rectangular amplitudes
own the same value at $t=0$.
These features are what we expect~\cite{Sharpe:1992pp}.

The vacuum amplitude is quite small up until $t \sim 8-14$,
and loss of the signals after that.
This characteristic is in agreement with the Okubo-Zweig-Iizuka (OZI) rule
and $\chi$PT in leading order, which predicts the vanishing
of the vacuum amplitude~\cite{Kuramashi:1993ka}.

\begin{figure}[th]
\begin{center}
\includegraphics[width=8.0cm]{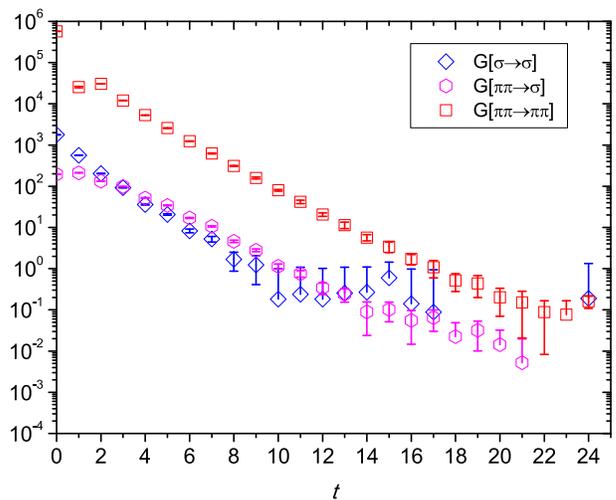}
\end{center}
\vspace{-0.5cm}
\caption{ \label{fig:G_t}
Real parts of the diagonal components
($\pi\pi\to\pi\pi$ and $\sigma\to\sigma$)
and the off-diagonal component $\pi\pi\to\sigma$.
Occasional points with negative central values
for the off-diagonal component $\pi\pi\to\sigma$ and the diagonal
component $\sigma \to \sigma$ are not displayed.
}
\end{figure}

In figure~\ref{fig:G_t}, we display the real parts
of the diagonal components
($\pi\pi\to\pi\pi$ and $\sigma\to\sigma$)
and the off-diagonal component $\pi\pi\to\sigma$
for the correlation function $C(t)$.
As we discussed in~\cite{Fu:2011zzh,Bernard:2007qf,Fu:2011zzl},
there exists the bubble contribution in sigma correlator,
thus we will compute the scattering phase shift
with the bubble term deducted from the sigma correlator.
In Refs.~\cite{Bernard:2007qf,Fu:2011zzl},
we parameterized the bubble term
by three low-energy coupling constants which
were fixed to our previous determined
values~\cite{Fu:2011zzh,Fu:2011zzl} in our concrete calculation.
After removing the bubble term, the remaining sigma correlator
has the clean information for sigma meson.

\begin{figure}[th]
\begin{center}
\includegraphics[width=8.0cm]{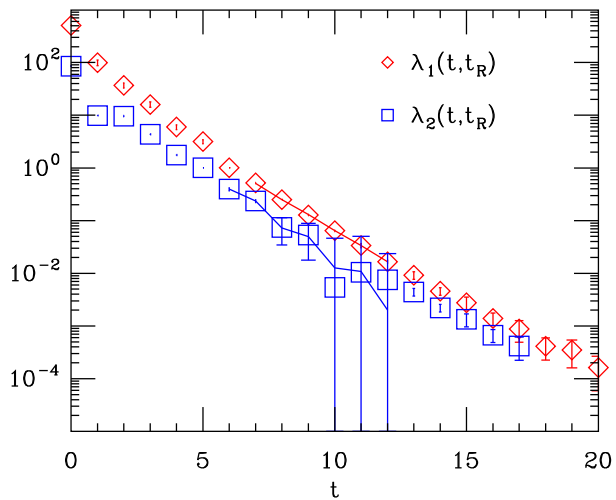}
\end{center}
\vspace{-0.5cm}
\caption{\label{fig:Lambda_t}
The eigenvalues $\lambda_1(t,t_R)$ and $\lambda_2(t, t_R)$
as the function of $t$.
The solid lines are correlated fits to eq.~(\ref{Eq:asy}).
Occasional points with negative values in $\lambda_2(t, t_R)$ are not shown.
}
\end{figure}

We calculate two eigenvalues $\lambda_n(t,t_R)$ ($n=1,2$)
for the matrix $M(t,t_R)$ in eq.~(\ref{eq:M_def})
with the reference time $t_R$.
In figure~\ref{fig:Lambda_t},
we plot our lattice results for $\lambda_n(t, t_R) (n = 1, 2)$
as a function of time $t$ together with a correlated fit using eq.~(\ref{Eq:asy}).
From these fits we can extract the energies
which will be used to obtain the scattering phase.

To achieve the energies reliably,
we should take two systematic errors into considerations:
the excited states and the warp-around effects~\cite{Feng:2010es}.
By denoting a fitting range $[t_{\mathrm{min}}, t_{\mathrm{max}}]$
and changing $t_{\mathrm{min}}$ values and $t_{\mathrm{max}}$ numbers,
we can suppress these systematic errors.
In our concrete calculation, we take $t_\mathrm{min}=t_R+1$ and
increase $t_R$ to restrain the excited state contributions~\cite{Feng:2010es}.
Moreover, we select $t_\mathrm{max}$ to be enough
aloof from the time slice $T/2$ to avert the warp-around contributions~\cite{Feng:2010es}.
The fitting parameters $t_R$, $t_{\mathrm{min}}$ and $t_{\mathrm{max}}$,
fit quality $\chi^2/\mathrm{dof}$ together with the fit results for
$\overline{E}_n$ ($n=1,2$)
are summarized in table~\ref{tab:fitting_results}.
\begin{table}[h!]
\centering
\caption{\label{tab:fitting_results}
The values of the energy eigenvalues for the ground
state ($n=1$) and first excited state ($n=2$).
In table we list the reference time $t_R$,
fitting range: $t_{\mathrm{min}}$ and $t_{\mathrm{max}}$,
fit quality $\chi^2/\mathrm{dof}$
and fit values for $\overline{E}_n$ ($n=1,2$) in lattice units.
}
\begin{ruledtabular}
\begin{tabular}{cccccc}
n&$t_R$&$t_{\mathrm{min}}$&$t_{\mathrm{max}}$
& $a\overline{E}_n$ &$\chi^2/\mathrm{dof}$ \\
\hline
$1$  &  $6$  &  $7$ & $12$  & $0.6767(20)$  & $0.922/2$ \\
$2$  &  $5$  &  $6$ & $12$  & $0.8086(75)$  & $0.158/3$ \\
\end{tabular}
\end{ruledtabular}
\end{table}

The mass $m_\pi$ and energy $E_\pi$ are achieved by a one-pole fit to
$C_\pi (t;{\mathbf 0})$ and $C_\pi(t;{\mathbf p})$
in eq.~(\ref{eq:Gpi}), respectively.
Then the energy of the free pions $E_1^0$
take as $E_1^0 = m_\pi + E_\pi$.
These results are summarized in table~\ref{table:TanDel} in lattice units.
We note that $\overline{E}_1 < E_1^0 < \overline{E}_2$,
which indicates a resonance existing in between.
\begin{table}[th]
\centering
\caption{ \label{table:TanDel}
Summary of the energy $\overline{E}_n$
and the scattering phase shift $\delta$.
The invariant mass $\sqrt{s}$,
the momentum $k$ and the phase shifts $\delta$
calculated with eq.~(\ref{eq:Disp_Two_Cont_k}) are referred to as {\it Cont},
and those obtained with eq.~(\ref{eq:Disp_Two_Lat_k}) are referred to as {\it Lat}.
The scattering momentum $k_0$ is calculated by $k_0^2 = s/4  - m_\pi^2$.
}
\begin{ruledtabular}
\begin{tabular}{ l l l l l }
       & \multicolumn{1}{c}{ $n=1$ } &
       & \multicolumn{1}{c}{ $n=2$ } &  \\
\hline
$E^0_n$          &  $ 0.7085(6) $  &
                 & \multicolumn{1}{c}{-----} &  \\
$\overline{E}_n$ &  $ 0.6767(20) $  &
                 &  $ 0.8086(38) $      \\
\hline
& \multicolumn{1}{c}{Cont} & \multicolumn{1}{c}{Lat}
& \multicolumn{1}{c}{Cont} & \multicolumn{1}{c}{Lat}  \\
$\sqrt{s}$       &  $ 0.5511(25)$    & $0.5613(25)$
                 &  $ 0.7068(43)$    & $0.7179(44)$  \\
$k^2$            &  $ 0.0155(7) $    & $0.0185(8)$
                 &  $ 0.0644(15)$    & $0.0699(17)$  \\
$k_0^2$  & \multicolumn{1}{c}{-----} & $0.0182(7)$
         & \multicolumn{1}{c}{-----} & $0.0684(16)$  \\
$\tan\delta$     &  $ 0.380(17)$     & $ 0.257(14)$
                 &  $-1.261(44)$     & $-1.509(55)$  \\
$\sin^2\delta$   &  $ 0.126(9)$     & $ 0.0621(64)$
                 &  $ 0.614(16)$     & $ 0.695(15) $  \\
\end{tabular}
\end{ruledtabular}
\end{table}

In table~\ref{table:Disp} we give the pion mass and its the energy
with the momentum ${\mathbf p}=(2\pi/L){\mathbf e}_3$,
calculated from the pion correlator.
Also we show the sigma mass and its the energy  with the same momentum,
calculated from the $\sigma$ correlator.
\begin{table}[th]
\centering
\caption{\label{table:Disp}
The mass $m$ of the $\pi$ and $\sigma$ meson,
and the energy $E$ of the $\pi$ and $\sigma$ meson
with momentum ${\mathbf p} = (2\pi/L) {\mathbf e}_3$ in lattice units.
}
\begin{ruledtabular}
\begin{tabular}{ c l c }
     &  $\pi$        &  $\sigma$     \\
\hline
$am$  &  $0.2459(2)$  &  $0.594(35)$  \\
$aE$  &  $0.4626(5)$  &  $0.714(22)$  \\
\end{tabular}
\end{ruledtabular}
\end{table}

\subsection{Lattice discretization effects }
\label{SubSec: Effect of finite lattice spacing }
We should premeditate the discretization error
in Rummukainen-Gottlieb formula (\ref{eq:Luscher_MF}).
It comes from the Lorentz transformation
from the MF to the CM frame.
In Lorentz transformation we use,
\begin{equation}
\sqrt{s} = \sqrt{ E_{MF}^2 - p^2 } , \quad
k^2      = \frac{s}{4} - m_\pi^2 .
\label{eq:Disp_Two_Cont_k}
\end{equation}
On the lattice, Rummukainen and Gottlieb~\cite{Rummukainen:1995vs}
suggest using the lattice modified relations,
\begin{eqnarray}
\cosh( \sqrt{s} ) &=& \cosh(E_{MF}) - 2\sin^2(p/2)  , \cr
2\sin^2 (k/2)     &=&
\cosh\left( \frac{\sqrt{s}}{2}\right)- \cosh(m_\pi) .
\label{eq:Disp_Two_Lat_k}
\end{eqnarray}
To comprehend the discretization effects,
we calculate invariant mass $\sqrt{s}$
and momentum $k$ from the relations
both in the continuum (\ref{eq:Disp_Two_Cont_k})
and on the lattice (\ref{eq:Disp_Two_Lat_k}),
and then calculate the phase shift.
We regard the difference stemming from
two choices as the discretization error.
The results for the invariant mass $\sqrt{s}$,
momentum $k$ and phase shift $\delta$
are tabled in table~\ref{table:TanDel} in lattice units.

\subsection{Extraction of resonance parameters}
\label{SubSec:Scattering Phase Shift and Decay Width }
\begin{figure}[th]
\begin{center}
\includegraphics[width=8.0cm]{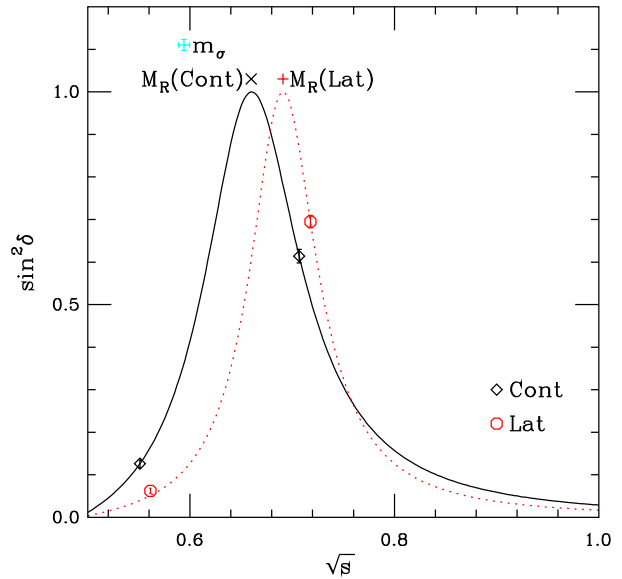}
\end{center}
\vspace{-0.5cm}
\caption{\label{fig:sin2Del}
Scattering phase shift $\sin^2\delta$, positions of $m_\sigma$
and resonance mass $M_R$ are displayed.
 {Cont} refer to the results achieved with (\ref{eq:Disp_Two_Cont_k})
and {Lat} to those with (\ref{eq:Disp_Two_Lat_k}).
The two lines are taken by eq.~(\ref{eq:tanDel_g})
with parameters $g_{\sigma\pi\pi}$ and $M_R$
obtained in eq.~(\ref{eq:FinalR_Cont}) and eq.~(\ref{eq:FinalR_Lat}), respectively.
}
\end{figure}

From table~\ref{table:TanDel},
the differences due to the choice of the energy-momentum relations
are obviously observed in $\sqrt{s}$ and $k$ .
Moreover, the difference for phase shift $\delta$ is
significantly larger than the statistical errors.
These are also shown in figure~\ref{fig:sin2Del},
where the phase shift $\sin^2 \delta$ is drawn
and the abscissa is in lattice units.
In table~\ref{table:TanDel} we see that
the sign of the phase shift $\delta$
at $\sqrt{s}< m_\sigma$ ($am_\sigma = 0.594(33)$) is positive,
and that at $\sqrt{s}> m_\sigma$ is negative.
These features confirm the presence of a resonance
around $\sigma$ mass.

In practice, we should extract the $\sigma$ meson decay width
by fitting the phase shift data with the RBWF
since the kinematic factors in the decay width
depend explicitly on the quark mass~\cite{Nebreda:2010wv}.
However, in this work,  we just measured a lattice data
on a set of quark mass,
so we must take an alternative approach.
As we discussed in section~\ref{SubSec:Scattering_Phase},
we parameterize the resonant characteristic of
the $\delta_0$ using the effective 
$\sigma\to\pi\pi$ coupling constant $g_{\sigma\pi\pi}$, namely,
\begin{equation}
\tan\delta_0 = \frac{g_{\sigma\pi\pi}^2}{8\pi}\frac{k}{\sqrt{s}(M_R^2-s)} ,
\label{eq:tanDel_g}
\end{equation}
where $M_R$ is the resonance mass.

According to the discussions in ref.~\cite{Nebreda:2010wv},
we can reasonably assume that the coupling constant $g_{\sigma\pi\pi}$ is a constant since it changes quite slowly as the quark mass varies.
Thus, equation~(\ref{eq:tanDel_g}) allows us
to solve for two unknown parameters:
the coupling constant $g_{\sigma\pi\pi}$,
and resonance mass $m_R$.
The discretization error may arise from the choice of
$\sqrt{s}$ and $k$.
Fortunately, our lattice  results show that
this does not cause a serious problem numerically.
In table~\ref{table:TanDel}
we present the momentum $k_0$ evaluated by $k_0^2 = s/4 - m_\pi^2$.
We notice that the difference between $k$ and $k_0$ is not considerable.
Thus, we can ignore this systemic error for the current study.
Actually, we use the momentum $k_0$ when applying eq.~(\ref{eq:tanDel_g}).

The coupling constant $g_{\sigma\pi\pi}$
and the resonance mass $M_R$ solved by eq.~(\ref{eq:tanDel_g}) read
\begin{eqnarray}
g_{\sigma\pi\pi} &=&  3.22(52)  \, {\rm GeV} ,  \cr
M_R              &=&  0.660(31)              ,  \cr
M_R / m_\sigma   &=&  1.112(85)              ,
\label{eq:FinalR_Cont}
\end{eqnarray}
where we utilize the eq.~(\ref{eq:Disp_Two_Cont_k}).
If we employ the eq.~(\ref{eq:Disp_Two_Lat_k}),
we achieve
\begin{eqnarray}
g_{\sigma\pi\pi} &=&  2.69(44) \, {\rm GeV} ,   \cr
M_R              &=&  0.691(37)             ,   \cr
M_R / m_\sigma   &=&  1.163(93)             .
\label{eq:FinalR_Lat}
\end{eqnarray}
The value of the coupling constant $g_{\sigma\pi\pi}$ is
in reasonable agreement with $g_{\sigma\pi\pi} = 2.47(45) \, {\rm GeV}$
obtained in ref.~\cite{Kaminski:2009qg},
$g_{\sigma\pi\pi} = 2.97(4)  \, {\rm GeV}$~\cite{Oller:2003vf}
and $g_{\sigma\pi\pi} = 2.86 \, {\rm GeV}$~\cite{Nebreda:2010wv}.

In figure~\ref{fig:sin2Del}, we display the curves for $\sin^2\delta_0$
solved by eq.~(\ref{eq:tanDel_g})
with the coupling constant $g_{\sigma\pi\pi}$ and
the resonance mass $M_R$ given in eq.~(\ref{eq:FinalR_Cont})
and eq.~(\ref{eq:FinalR_Lat}), respectively.
The position of the resonance mass $M_R$ (at $\sin^2\delta_0=1$)
are also displayed in figure~\ref{fig:sin2Del} for two cases
(black cross and red plus for the continuum and lattice cases, respectively).
For visualized comparison,
we also draw the sigma mass $m_\sigma$ (fancy cyan plus),
which is in reasonable agreement with
the $m_R$.

Supposing that the quark dependence of $g_{\sigma\pi\pi}$
is quite small~\cite{Nebreda:2010wv},
we can roughly calculate $\sigma$ meson decay width
at the physical point as
$$
\Gamma^{\rm phy} = \frac{g_{\sigma\pi\pi}^2}{8\pi}
\frac{k^{\rm phy}}{(m_\sigma^{\rm phy})^2} ,
$$
where $m_\sigma^{\rm phy}=513\pm32$ MeV is
the estimated physical $\sigma$ meson mass taken from PDG~\cite{Nakamura:2010zzi},
and momentum $k^{\rm phy}$ is calculated by
$$
(k^{\rm phy})^2 = \frac{ (m_\sigma^{\rm phy})^2}{4} - (m_\pi^{\rm phy})^2  ,
$$
where $m_\pi^{\rm phy}$ is physical pion mass
($m_\pi^{\rm phy} \approx 140$ MeV)~\cite{Nakamura:2010zzi}.
This produces
\begin{equation}
\label{eq:FinalR_Gamm_Cont}
\Gamma^{\rm phy} = (337 \pm 82) \, {\rm MeV} \,
\end{equation}
where we use the data given in eq.~(\ref{eq:FinalR_Cont}), and
\begin{equation}
\Gamma^{\rm phy} = (236 \pm 49) \, {\rm MeV}
\label{eq:FinalR_Gamm_Lat}
\end{equation}
where we employ the data given in eq.~(\ref{eq:FinalR_Lat}).
We can observe that the difference stemming from
two choices of the energy-momentum relations
is larger than with the statistical error.
Although our preliminary estimates for the $\sigma \to \pi\pi$ decay width
in this work is not within the PDG
estimated result $\Gamma = 600-1000 $MeV~\cite{Nakamura:2010zzi},
this is still an inspiring result,
considering that we make a big assumption
about the coupling constant does not depend on the quark mass,
an perform a long chiral extrapolation, etc.

In the present study, we make an extensive use of the RBWF.
It is well-known that the sigma meson is a very wide object and the
RBWF approximation holds perfectly for relatively narrower objects.
As discussed in ref.~\cite{Doring:2011vk},
we should adopt a much more model-independent approach to
the extraction of the finite volume limit.
In refs.~\cite{Doring:2011nd,Roca:2012rx,Chen:2012rp},
Oset et al. pointed out that if we have got three energies in the cubic box,
with the momentum $p=0$ and $p$ different of zero,
we can still use the finite volume formulas to get the phase shifts in a correct manner. Alternative methods are also discussed in these references.
In our future tasks, we must address the phenomenological treatment.

\section{Conclusions}
\label{Sec:Conclusions}
In this work, we have carried out a lattice calculation
of the $s$-wave $\pi\pi$ scattering phase shift for isospin $I = 0$ channel
near $\sigma$-meson resonance with total non-zero momentum in one MF,
for MILC ``medium'' coarse ($a=0.15$ fm) lattice ensemble.
We employed the technique in refs.~\cite{Kuramashi:1993ka,Fukugita:1994ve},
namely, the moving wall source
without gauge fixing for the $I = 0$ channel
to obtain the reliable precision.

We have demonstrated that
the phase shift data clearly shows the presence of
a resonance at a mass around $\sigma$ meson mass.
Moreover, we extracted $\sigma$ meson decay width
from the phase shift data and showed that
it is fairly compared with the corresponding PDG estimation~\cite{Nakamura:2010zzi}.

We adopted the ERF, which allows
us to use the effective $\sigma \to \pi\pi$ coupling constant
$g_{\sigma\pi\pi}$ to extrapolate
our lattice simulation point $m_\pi/m_\sigma \approx 0.414$
to the physical point $m_\pi/m_\sigma \approx 0.273$,
assuming that $g_{\sigma\pi\pi}$ is independent of quark mass.
This is just a rough estimation. 
We are planning to improve it.

When our preliminary lattice results reported here
are compared with its PDG quantities,
it is clear that the lattice simulations is just rough estimation,
and even can not be considered to be``physical'' one.
So we view our rudimentary works presented here as
stepping out a first step to
the study of $\sigma$ resonance from lattice QCD.

\section*{Acknowledgments}
We are grateful to MILC Collaboration
for using Asqtad lattice ensemble and MILC codes.
We should thank Eulogio Oset
for their encouraging and critical comments.
The computations for this work were carried out at AMAX,
CENTOS and HP workstations
in the Institute of Nuclear Science and Technology, Sichuan University.


%
\end{document}